\documentclass[letterpaper,twocolumn,10pt]{article}
\usepackage{usenix2019_v3}

\microtypecontext{spacing=nonfrench}

\usepackage{booktabs} 
\usepackage{pgf}
\usepackage{pgfplots}
\usepackage{textcomp}
\usepackage{pgfplotstable}
\pgfplotsset{compat=newest}
\colorlet{negro}{black}
\colorlet{gris}{black!70}
\colorlet{rojo}{red!70!black}
\colorlet{rojol}{red}
\usepackage{doi}

\usepackage{xspace} 
\usepackage{tikz}
\usepackage{tikz-uml}
\usetikzlibrary{petri} 
\usetikzlibrary{backgrounds}
\usetikzlibrary{calc}
\usepackage{xcolor}

\usepackage{algorithm}
\usepackage{algorithmic}
\usepackage{pgf-umlsd}
\usepackage{environ}

\usepackage{subcaption}

\usepackage[inline]{enumitem}
\usepackage{placeins}

\newcommand{\toolAbrev}{DIMAQS\xspace}
\newcommand{\toolLong}{DIMAQS (Dynamic Identification of Malicious Query Sequences)\xspace}

\newcommand{\controller}{\texttt{Controller}\xspace}
\newcommand{\Controller}{\texttt{Controller}\xspace}

\newcommand{\notifier}{\texttt{Notifier}\xspace}

\newcommand{\classifier}{\texttt{Classifier}\xspace}

\newcommand{\policyDB}{\texttt{Security Policy}\xspace}
\newcommand{\policy}{\texttt{Security Policy}\xspace}

\newcommand{\monitoring}{\texttt{Monitoring}\xspace}

\newcommand{\resolution}{\texttt{Incident Resolution}\xspace}
\newcommand{\Resolution}{\texttt{Incident Resolution}\xspace}

\newcommand{\rewriter}{\texttt{Query Rewriter}\xspace}
\newcommand{\Rewriter}{\texttt{Query Rewriter}\xspace}

\newcommand{\add}[1]{#1}

\newcommand{\libPetri}{libPetri\xspace}

\usepackage{cleveref}
\usepackage[colorinlistoftodos,prependcaption]{todonotes}
\usepackage{xargs}                      

\newcommandx{\todoLukas}[2][1=]{\todo[linecolor=red,inline,backgroundcolor=red!25,bordercolor=red,#1]{\textbf{Lukas: }#2}}
\newcommandx{\todoChristoph}[2][1=]{\todo[linecolor=blue,inline,backgroundcolor=blue!25,bordercolor=blue,#1]{\textbf{Christoph: }#2}}
\newcommandx{\todoMichael}[2][1=]{\todo[linecolor=green,inline,backgroundcolor=green!25,bordercolor=green,#1]{\textbf{Michael: }#2}}
\newcommandx{\todoAlexandra}[2][1=]{\todo[linecolor=yellow,inline,backgroundcolor=yellow!25,bordercolor=yellow,#1]{\textbf{Alexandra: }#2}}
\newcommandx{\todoAll}[2][1=]{\todo[linecolor=black,inline,backgroundcolor=black!25,bordercolor=black,#1]{\textbf{Everybody: }#2}}

\begin{document}


\title{\Large \bf Hands Off my Database: Ransomware Detection in Databases through Dynamic Analysis of Query Sequences}

\author{
{\rm Lukas Iffl\"ander}\\
University of W\"urzburg 
\and
{\rm Alexandra Dmitrienko}\\
University of W\"urzburg
\and
{\rm Christoph Hagen}\\
University of W\"urzburg
\and
{\rm Michael Jobst}\\
University of W\"urzburg
\and
{\rm Samuel Kounev}\\
University of W\"urzburg
} 

\maketitle


\begin{abstract}
Ransomware is an emerging threat which imposed a \$ 5 billion loss in 2017 and is predicted to hit 11.5 billion in 2019.  While initially targeting PC (client) platforms, ransomware recently made the leap to server-side databases -- starting in January 2017 with the MongoDB Apocalypse attack, followed by other attack waves targeting a wide range of DB types such as MongoDB, MySQL, ElasticSearch, Cassandra, Hadoop, and CouchDB. While previous research has developed countermeasures against client-side ransomware (e.g., CryptoDrop and ShieldFS), the problem of server-side ransomware has received zero attention so far.

\sloppy
In our work, we aim to bridge this gap and present \toolLong, a novel anti-ransomware solution for databases. 
\toolAbrev performs runtime monitoring of incoming queries and pattern matching using Colored Petri Nets (CPNs) for attack detection. Our system design exhibits several novel techniques to enable efficient detection of malicious query sequences globally (i.e., without limiting detection to distinct user connections). Our proof-of-concept implementation targets MySQL servers. 
The evaluation shows high efficiency with no false positives and no false negatives and very moderate performance overhead of under 5\%. We will publish our data sets and implementation allowing the community to reproduce our tests and compare to our results.
\end{abstract}

\section{Introduction}\label{sec:introduction}


In today's era of digital transformation, data has become more critical than ever before. The amount of data we produce daily is astonishing -- every day hundreds of millions of people are taking photos, make videos and exchange messages. Furthermore, data is not only an asset for users nowadays, but has also become the key component of digitization and transformation of today's businesses globally -- enterprises collect data on consumer preferences, purchases, and trends and use it to optimize their business models and strategies. Given such trends, the importance of database security is hard to overestimate -- the rapid growth of the data volume stored in the databases of service providers, in cloud environments and enterprise data centers, as well as their increasing importance, make them attractive attack targets.  




Traditionally, attacks on data have aimed to undermine confidentiality and authenticity. More recently, however, attacks against the availability  of data, services, and users have become common as well -- modern attackers deploy ransomware, malicious software that encrypts data and holds the decryption key until the victim pays a ransom. 
They still claim the ransom pretending to have encrypted the data.
The financial loss from ransomware is significant -- it reached 5 billion USD in 2017 and is predicted to hit 11.5 billion by 2019~\cite{RansomwareStats}. 

\vspace{0.2cm} \noindent\textbf{The rise of server-side ransomware} While the first ransomware attacks targeted client platforms (information stored in users' files), recently such attacks made a leap to server-side databases that store, accumulate and process (big) data.
In January 2017 tens of thousands of MongoDB servers were hit in an attack called MongoDB Apocalypse~\cite{MongoDB-apocalypse-2017,MongoDBattacks2017}, followed by a second attack wave targeting MySQL servers~\cite{GuardiCore17}. Since then, server-side ransomware attacks spread to a wide range of server technologies, including ElasticSearch~\cite{ElasticSearch-2017}, Cassandra~\cite{Cassandra2017}, Hadoop and CouchDB~\cite{Hadoop2017}. 

\vspace{0.2cm} \noindent\textbf{Attack scenario} The typical attack scenario of server-side ransomware observed so far is as follows: First, an attacker gains remote privileged access to the database database through the exploitation of configuration vulnerabilities such as the usage of default passwords
\footnote{Note that default passwords and other misconfiguration errors are prevalent real-world problems. For instance, Mirai botnet\cite{Ant2017} used similar vulnerabilities to take over more than 600,000 IoT devices arond the globe.}. Once connected, they execute commands for data enumeration (e.g., to learn names of databases and tables hosted), then drop (delete) data and insert the ransom message with instructions how to pay the ransom. Remarkably, in contrast to client-side ransomware, the new attack form wipes the data without making any plaintext or encrypted copy, e.g., acting as a \emph{wiper}. This strategy has, on the one hand, more dramatic implications for the victim, since the data is unrecoverable even if the ransom is paid. On the other hand, the attack is stealthier, since no intensive and easily detectable operations required, such as bulk encryption or massive data copying, and no back channel to the attacker needed (e.g., for delivering the decryption key or recovered data) that could be used to trace them back.


\vspace{0.2cm}\noindent\textbf{Motivation for server-side ransomware to spread} While server-side ransomware is more recent and to this day less widespread than client-side ransomware, there are reasons why the situation might change quite soon. First, enterprises can afford to pay higher ransoms than private users. As a comparison, the typical ransom amount for regular users lies in the range of a few hundred dollars. However, businesses can pay much more -- for instance, in a recent attack, a Los Angeles Hospital paid USD 17\,000 of ransom to attackers~\cite{LosAngelesHospital}. 
Second, in recent years, researchers and antivirus companies developed countermeasures against client-side ransomware. However, to date, no solutions exist against ransomware targeting database servers. This lack of protection makes databases easy attack targets. 

\vspace{0.2cm} \noindent\textbf{Do victims pay the ransom to a wiper?} Note that there is evidence that even though server-side ransomware is a wiper, some desperate victims paid the ransom, nonetheless. We identified that two known ransomware addresses involved in MySQL attacks~\cite{GuardiCore17} received 0.6 BTC (equivalent to 3 payments). For the attacks against MongoDB, we identified a total of 160 ransom payments to the addresses collected in~\cite{MongoDBattacks2017}, totaling in 26.35 BTC. Moreover, the survey~\cite{MongoDBattacks2017} reveals that even production systems lack sufficient protection by strong passwords and sensible backup strategy: Among 123 surveyed ransom victims, only 11\% had recent backups, and 8\% paid the ransom. 


\vspace{0.2cm}\noindent\textbf{State of the art}
Existing anti-ransomware solutions are aiming at detection of client-side ransomware only. They follow two dominant strategies: Signature-based detection of malicious binaries and runtime monitoring and behavioral analysis for anomaly detection.
The first one builds upon detection of malicious binaries and is typically used by anti-virus vendors, while the second strategy originates from research papers~\cite{continella2016shieldfs,continella2017shieldfs,scaife2016cryptolock,kolodenker2017pay} and relies on runtime monitoring of file accesses and the detection of malicious activity based on heuristics, such as access to multiple files, their modification, and renaming. Unfortunately, both strategies are not applicable for detection of database wipers. Since in server-side ransomware attack scenario an attacker connects to the database remotely, there is no malicious binary on the platform that could be detected. Furthermore, monitoring at the file system level for abnormal activity is not adequate either since there is no direct correlation between an attacker's activity and file access patterns.

\vspace{0.2cm}
\noindent\textbf{Our contributions} In this paper, we aim to improve the security of database systems and propose \toolLong, signature-based intrusion detection tool that can detect sequences of malicious queries. Generally, the tool is not limited to ransomware detection and can potentially be applied to detection of other attack classes as long as they rely on malicious  sequences of queries (e.g., advanced SQL injections aiming at removing code execution~\cite{BlackHat2009-SQL-advanced}). However, motivated by the rise of server-side ransomware we apply it to the problem of ransomware detection. We make the following contributions:



\begin{itemize}[leftmargin=7pt]
    \item 
    We provide design and implementation of \toolAbrev, a framework that can detect sequences of malicious queries.  To keep track of queries and to perform detection, our solution leverages Colored Petri Nets (CPNs) to model the series of events used in attacks and to match them to known malicious patterns. Our system design exhibits several novel techniques (dynamic creation of colors, merging of tokens and token expiration) to reduce the complexity of the system representation and achieve better performance. Our framework performs system-wide monitoring and as such can detect malicious sequences injected through several user sessions and interleaved with benign queries -- a quite interesting feature that eliminates most obvious evasion strategies. Our implementation targets MySQL, one of the most popular database management systems, and imposes only a very moderate performance overhead under 5\%. We realize our solution in the form of a MySQL plugin that is easily installable on existing MySQL servers, thus preserving compatibility with legacy software. We will publish the source code on GitHub along with the paper.

    \item We apply \toolAbrev to the challenging problem of server-side ransomware. To make detection of such attacks possible, we analyze previously observed attacks and extract their distinctive properties that provide a basis for attack detection. We then evaluate the effectiveness and practicality of our solution using three data sets: Malicious data set recorded by us, and benign query sets from a publication management system and a MediaWiki server. The results demonstrate the high efficiency of our approach with no false negatives or false positives. We will publish our data sets along with the paper to the benefit of the research community. To the best of our knowledge, our malicious data set will be the first one publicly available.

    
\end{itemize}


\begin{figure*}[!htb]
    \centering
    \begin{subfigure}[c]{0.3\textwidth}

\centering

\begin{tikzpicture}[yscale=-1,thin,>=stealth,
every transition/.style={fill,minimum width=1mm,minimum height=3.5mm},
every place/.style={draw,thick,minimum size=8mm}]
    \node[place,label=above:$p_1$,tokens=1] (p1) at (0,0) {};
    \node[place,label=below:$p_2$,tokens=2] (p2) at (0,1) {};
    \node[place,label=below:$p_3$] (p3) at (4,0.5) {};
    \node[transition, label=above:$t_1$] at (2,0.5) {}   
        edge [pre] node[above] {2} (p1)
        edge [pre] node[below] {1} (p2)
        edge [post] node[auto] {1} (p3);
\end{tikzpicture}
\subcaption{Transition disabled}
\label{subfig:background_pn_disabled}

\end{subfigure}
\begin{subfigure}[c]{0.3\textwidth}

\centering

\begin{tikzpicture}[yscale=-1,thin,>=stealth,
every transition/.style={fill,minimum width=1mm,minimum height=3.5mm},
every place/.style={draw,thick,minimum size=8mm}]
    \node[place,label=above:$p_1$,tokens=2] (p1) at (0,0) {};
    \node[place,label=below:$p_2$,tokens=2] (p2) at (0,1) {};
    \node[place,label=below:$p_3$] (p3) at (4,0.5) {};
    \node[transition, label=above:$t_1$] at (2,0.5) {}   
        edge [pre] node[above] {2} (p1)
        edge [pre] node[below] {1} (p2)
        edge [post] node[auto] {1} (p3);
\end{tikzpicture}
\subcaption{Transition enabled}
\label{subfig:background_pn_enabled}

\end{subfigure}
\begin{subfigure}[c]{0.3\textwidth}

\centering

\begin{tikzpicture}[yscale=-1,thin,>=stealth,
every transition/.style={fill,minimum width=1mm,minimum height=3.5mm},
every place/.style={draw,thick,minimum size=8mm}]
    \node[place,label=above:$p_1$,tokens=0] (p1) at (0,0) {};
    \node[place,label=below:$p_2$,tokens=1] (p2) at (0,1) {};
    \node[place,label=below:$p_3$,tokens=1] (p3) at (4,0.5) {};
    \node[transition, label=above:$t_1$] at (2,0.5) {}   
        edge [pre] node[above] {2} (p1)
        edge [pre] node[below] {1} (p2)
        edge [post] node[auto] {1} (p3);
\end{tikzpicture}
\subcaption{Transition fired}
\label{subfig:background_pn_fired}

\end{subfigure}
    \caption{Demonstration of Petri net execution using a simple example}
    \label{fig:petri_net_example}
    \centering
    \begin{subfigure}[c]{0.3\textwidth}

\centering

\begin{tikzpicture}[yscale=-1.1,thin,>=stealth,
every transition/.style={fill,minimum width=1mm,minimum height=3.5mm},
every place/.style={draw,thick,minimum size=8mm}]
    \node[place,label=above:$p_1$] (p1) at (0,0) {}
        [children are tokens]
        child {node [token,fill=red] {2}}
        child {node [token] {1}}
        child {node [token,fill=red] {2}};
    \node[place,label=below:$p_3$] (p3) at (4,0) {};
    \node[transition, label=above:$t_1$] at (2,0) {}   
        edge [pre] node[auto] {(2,1)} (p1)
        edge [post] node[auto] {(1,0)} (p3);
\end{tikzpicture}
\subcaption{Transition disabled}
\label{subfig:background_cpn_disabled}

\end{subfigure}
\begin{subfigure}[c]{0.3\textwidth}

\centering

\begin{tikzpicture}[yscale=-1.1,thin,>=stealth,
every transition/.style={fill,minimum width=1mm,minimum height=3.5mm},
every place/.style={draw,thick,minimum size=8mm}]
    \node[place,label=above:$p_1$] (p1) at (0,0) {}
        [children are tokens]
        child {node [token,fill=red] {2}}
        child {node [token,fill=red] {2}}
        child {node [token] {1}}
        child {node [token] {1}};
    \node[place,label=below:$p_3$] (p3) at (4,0) {};
    \node[transition, label=above:$t_1$] at (2,0) {}   
        edge [pre] node[auto] {(2,1)} (p1)
        edge [post] node[auto] {(1,0)} (p3);
\end{tikzpicture}
\subcaption{Transition enabled}
\label{subfig:background_cpn_enabled}

\end{subfigure}
\begin{subfigure}[c]{0.3\textwidth}

\centering

\begin{tikzpicture}[yscale=-1.1,thin,>=stealth,
every transition/.style={fill,minimum width=1mm,minimum height=3.5mm},
every place/.style={draw,thick,minimum size=8mm}]
    \node[place,label=above:$p_1$] (p1) at (0,0) {}
        [children are tokens]
        child {node [token,fill=red] {2}};
    \node[place,label=below:$p_3$] (p3) at (4,0) {}
        [children are tokens]
        child {node [token] {1}};
    \node[transition, label=above:$t_1$] at (2,0) {}   
        edge [pre] node[auto] {(2,1)} (p1)
        edge [post] node[auto] {(1,0)} (p3);
\end{tikzpicture}
\subcaption{Transition fired}
\label{subfig:background_cpn_fired}

\end{subfigure}
    \caption{Colored Petri Net example. In comparison to the regular Petri net depicted in Figure~\ref{fig:petri_net_example}, the number of required \emph{places} is reduced from two to one without reducing functionality.}
    \label{fig:colored_petri_net_example}
\end{figure*}

\vspace{0.1cm}
\noindent
\textbf{Outline.}  The remainder of this paper is structured as follows. 
In \Cref{sec:background} we present the necessary background followed by system design of \toolAbrev in \Cref{sec:approach}. 
In \Cref{sec:implementation}, we reveal the details of our prototype implementation. Prototype evaluation results are presented in \Cref{sec:evaluation}. 
After a review of the related work in \Cref{sec:related}, we conclude the paper and outline future work in \Cref{sec:conclusion}.

\section{Background}\label{sec:background}

In this section, we provide the necessary background on Petri nets and their enhanced version, colored Petri nets. 

\vspace{0.2cm}
\noindent \textbf{Petri Nets}
are a commonly used mathematical modeling language for the description of distributed systems~\cite{peterson1981petri} named after their inventor Carl Adam Petri. They are a class of discrete event dynamic systems. 
A Petri net is a directed bipartite graph, in which nodes represent \emph{places} and \emph{transitions}, while edges, called \emph{arcs}, connect either a place to a transition or a transition to a place, but never connect two places or two transitions directly. Transitions are events in the system, and places are conditions that need to be satisfied for the transition to fire. 


Places may contain a discrete number of marks called \emph{tokens}.
Transitions \emph{fire} if they are \emph{enabled}, which is achievable by placing enough input tokens on the input places -- i.e., places directly connected to the transition. 
The value of the arc defines the number of tokens required per place. 
Once a transition fires, it consumes the required number of input tokens from the input places. 
The transition results in creating the specified number of output tokens on the places with arcs from the transition to them (output places).

Figure~\ref{fig:petri_net_example} shows a simple example of a Petri net.
The depicted Petri net consists of three places (depicted as circles), one transition (depicted as a bar), and three arcs.
Enabling the transition requires three tokens:  
Two tokens at place $p_1$ and one token at place $p_2$.
In Figure~\ref{subfig:background_pn_disabled} only one token is available at $p_1$. 
Regardless of the total count being three tokens, with only one token on $p_1$, the transition is not yet enabled.
Adding another token to $p_1$ in Figure~\ref{subfig:background_pn_enabled} satisfies the requirement and thus enables the transition.
When the transition fires, two tokens are subtracted from the token set at $p_1$ as well as one token from $p_2$. 
At the same time, the transition adds one token to $p_3$.
Figure \ref{subfig:background_pn_fired} shows the state after the transition firing.

Petri nets are a powerful tool for modeling~\cite{Chen2005Apr} and allow for extensions to suit various tasks like queuing Petri nets for performance modeling. 
In this work, we use colored Petri nets, an extension to ordinary Petri nets. 

\vspace{0.2cm}
\noindent \textbf{Colored Petri Nets} (CPNs) enable support for tokens of different types, also known as \emph{token colors}. 
Places can now contain tokens of multiple colors.
Arcs can define any combination of the colors for the number of input and output tokens.
This addition allows for making Petri nets more compact.

Figure~\ref{fig:colored_petri_net_example} illustrates the reduction in representation complexity by presenting a CPN derived from the previous example. 
The places $p_1$ and $p_2$ depicted in Figure~\ref{fig:petri_net_example} are now merged into a single place denoted as $p_1$, while tokens are now assigned different colors:    
Tokens formerly placed in $p_1$ are now black (1) and those placed in $p_2$ are red (2).
The transition now requires two black and one red token instead of requiring two tokens from $p_1$ and one from $p_2$. 
The overall Figure~\ref{fig:colored_petri_net_example} depicts the same process as before.
In Figure~\ref{subfig:background_cpn_disabled}, one black token is missing for the transition to be enabled.
In Figure~\ref{subfig:background_cpn_enabled} this token is added, thus enabling the transition.
Finally, in Figure~\ref{subfig:background_cpn_fired} the transition has fired, subtracting two black and one red token from $p_1$ and adding a black token to $p_3$.


\section{Design}\label{sec:approach}


\toolAbrev is the first system that aims at the detection of ransomware attacks in databases. 
In a nutshell, it represents an intrusion detection system that leverages knowledge about the attack pattern (or signature) and performs real-time system monitoring and pattern matching to detect intrusion attempts. For pattern matching, we leverage a CPN to encode the system states and their transitions inside the color information to detect when the system transitions to the state associated with the attack description.

The usage of (colored) Petri nets is a known technique for pattern matching, and their application to intrusion detection problems was investigated in previous works~\cite{kumar1994pattern,hu2003ident}. However, typical application scenarios of CPN-based intrusion detection systems target other environments, e.g., networks~\cite{verwoerd2002intrusion} and operating systems~\cite{axelsson2000intrusion}. 

The application of Petri nets for intrusion detection in databases was only considered by Hu et al.~\cite{hu2003ident}, who aimed at detection of anomalies of any sort, not specific to ransomware. However, they use \emph{uncolored} Petri nets and leverage them to model benign states of a database system rather than attack states. Hence, their solution requires a training phase to gain knowledge about the underlying data structure as well as about benign data update patterns. In contrast, our system does not require similar training. Moreover, their work is theoretical. Hence, they did not provide any implementation or evaluation results with which to compare. 


In our work, we aim to fill the gap and address the problem of ransomware attacks targeting databases. As such, we investigate the applicability of CPNs for ransomware attack detection in databases. We observe that databases are complex systems and modeling their state regarding dependency relationships and update patterns, as, e.g., done in~\cite{hu2003ident}, may lead to overly complicated system representations (for large and complex databases) and non-trivial overhead. Hence, we tackle the problem differently and choose to model malicious query sequences -- an approach which results in a much simpler system representation, and independence from the structure of the underlying data and update patterns. 

\add{Our approach is system-centric and allows for detection of attacks that are carried out over multiple sessions or multiple user accounts.}
We also develop several novel techniques that even further to simplify the system representation, namely 
(1) dynamic color creation (creating an infinite color space), (2) token merging and duplication, and (3) token expiration making the use of CPNs practical. 

The remaining part of this section is structured as follows: We first describe a typical ransomware attack scenario (Section~\ref{subsec:scenario}). Next, we present our adversary model (Section~\ref{subsec:advmodel}) followed by the system architecture description (Section~\ref{subsec:approach_architecture}). Finally, we show the interaction of the system components when handling incoming queries (Section~\ref{subsec:approach_beginMalicious}). 

\subsection{Attack Scenario} \label{subsec:scenario}



Our attack scenario originates from an analysis of a large-scale ransomware attack targeting MySQL servers that took place in February 2017~\cite{GuardiCore17}. The attacker performs the attack remotely by connecting to the database using a TCP connection. 
Once connected, an attacker gains root access through, e.g., brute-forcing the `root' password of the database. Next, they
enumerate the data in the database through retrieval of the list of the databases present. After that, the attacker creates a new table with an arbitrary name (e.g., the table with the name `WARNING'), either in a new database (e.g., named `PLEASE\_READ') or in an already existing database. This table includes a ransom message containing a contact email address as well as payment instructions to a bitcoin address. Finally, the attacker deletes (drops) the databases on the server and disconnects.

The scenario above describes the attack steps recorded in real-world attacks. Additionally, we accept that attack steps can deviate from this scenario: For instance, an attacker could first perform the database deletion and only after that insert the ransom message. Also, attackers may use arbitrary names for databases and tables and arbitrary patterns for the ransom message. We, however, assume that the attacker demands payments in cryptocurrency (such as Bitcoin or Ethereum) since they provide at least some level of anonymity in contrast to more traditional payment methods that involve banks\footnote{Since banks are obliged to follow "know your customer" policy.}. We also assume that an attacker continues to wipe data and does not aim to keep any data copies, since this would slow down the attack significantly, and would require storage on attacker's side and a communication channel between the victim and the attacker, which demands additional resources and increases chances of exposure. We also assume an attacker does not perform on-site database encryption since we did not identify any standard SQL commands that could be used to do so. 


\subsection{Adversary Model} \label{subsec:advmodel}

We make the following assumptions about the goal and the capabilities of the attacker. 
The attacker's goal is to destroy the available data and claim the ransom. 
We assume the remote attacker who is accessing the server over the Internet has no physical access to it. The software running on the server is trusted, i.e., the attacker has no malicious software installed on the system. However, the attacker has full access to the network and can communicate with the DBMS without any restrictions.
Furthermore, we assume an attacker with administrator-level privileges to the DBMS. This assumption is often fulfilled in practice  since the problem of weak or re-used passwords~\cite{ives2004} is well known and not satisfactory solved for over decades. 
For instance, findings show that most of the MySQL servers had no root password set due to using an insecure default configuration~\cite{MySQLMan}. Alternatively, an attacker might exploit a security vulnerability like~\cite{mySqlRootExploit} to gain administrator privileges for the database.

We, however, do not assume administrator privileges of the attacker to the operating system.
Also, we leave DoS attacks are out of our attacker model since an attacker with administrator privileges to DBMS can always cause a denial of service, e.g., through the creation of fake DBs or tables and exhausting DB's memory. 
\add{The attacker wants to perform a hit-and-run attack without considering other services and ways of communication.} 



\begin{figure}
    \centering
    \begin{tikzpicture}[every node/.style={rectangle,draw=black,minimum width=19mm,minimum height=2.5\baselineskip,inner sep=2mm}, >=stealth',very thick,black!50,text=black,every new ->/.style={shorten >=1pt}, graphs/every graph/.style={edges=rounded corners},node distance=0.6cm and 0.5 cm]
    \node (server) 		[fill=black!40!white]							{Server};
    \node (mysql_helper) [draw=none,left=of server] {};
    \node (rewriter_helper) [draw=none, right=of server] {};
	\node (mysql)		[draw=none,align=center,above left=-0.6 cm and -1.4 cm of mysql_helper]				{Database};
	\node (intermediate) [below= 0.1 cm of server, draw=none] {};
	\node (monitoring)		[left=of intermediate,align=center,fill=black!20!white]	    {Monitoring};
	\node (controller)	[below= 0.4 cm of intermediate, align=center] {Controller};
	\node (resolution)	[right=of controller, align=center]		    {Incident \\ Resolution};
	\node (parser) 		[right=of intermediate, align=center,fill=black!20!white]			{Query\\Rewriter};
	\node (notifier)	[below=of resolution]			    	{Notifier};
	\node (classifier)	[below=of controller, align=center]		{Classifier};
	\node (policyDB)	[left=of classifier, align=center]		{Security\\Policy};
	\node (tool_helper)		[draw=none,left=of controller]		 {};
	\node (tool) [draw=none, below= -0.1cm of policyDB] {\toolAbrev Plugin};
	\draw[->] (server) -| (monitoring) 
	    node[right,xshift=-0.5cm,draw=none,pos=0.7,font=\bfseries] {(1)};
	\draw[->] (monitoring) |- (controller)
	    node[left,xshift=0.5cm,draw=none,pos=0.3,font=\bfseries] {(2)};
	\draw[->] ($(controller.south west)!0.8!(controller.south)$) -- ($(classifier.north west)!0.8!(classifier.north)$)
		node[left,xshift=0.6cm,draw=none,pos=0.5,font=\bfseries] {(3)};
	\draw[->] (classifier) -- (policyDB)
	    node[above,draw=none,yshift=-0.2cm,pos=0.5,font=\bfseries] {(4)};
	\draw[->] ($(classifier.north east)!0.8!(classifier.north)$) -- ($(controller.south east)!0.8!(controller.south)$)
		node[right,xshift=-0.6cm,draw=none,pos=0.5,font=\bfseries] {(5)};
	\draw[->] ($(controller.north east)!0.7!(controller.east)$) -- ($(resolution.north west)!0.7!(resolution.west)$)
		node[above,draw=none,pos=0.5,yshift=-0.2cm,font=\bfseries] {(6)};
	\draw[->] (resolution) -- (parser)
		node[left,xshift=0.6cm,draw=none,pos=0.5,font=\bfseries] {(7)};
	\draw[->] (resolution.south) -- (notifier.north)
		node[left,xshift=0.6cm,draw=none,pos=0.5,font=\bfseries] {(8)};	
	\draw[->] ($(resolution.south west)!0.7!(resolution.west)$) -- ($(controller.south east)!0.7!(controller.east)$)
		node[below,draw=none,pos=0.5,yshift=0.2cm,font=\bfseries] {(9)};
	\draw[->] (controller) -- (server)
		node[left, xshift=0.6cm,draw=none,pos=0.6,font=\bfseries] {(10)};		
	\draw[->] ($(controller.west)!0.5!(controller.north west)$) -| ($(monitoring.south)!0.5!(monitoring.south east)$)
	    node[right,xshift=-0.6cm,draw=none,pos=0.7,font=\bfseries] {(11)};
	\begin{scope}[on background layer]
	    \node[draw,inner sep=3mm,fit=(monitoring.north west) (mysql.east) (rewriter_helper) (parser.north east), rounded corners, dashed] {};
	    \node[draw,inner sep=3mm,fit=(tool.east) (monitoring.south west) (parser.south east) (notifier), rounded corners, dashed] {};
	\end{scope}
\end{tikzpicture}

    \caption{System architecture of \toolAbrev. Dark grey boxes are components provided by the database, light grey boxes are components that interface between \toolAbrev and the database, and white boxes belong to \toolAbrev itself.}
    \label{fig:approach_dataflow}
\end{figure}
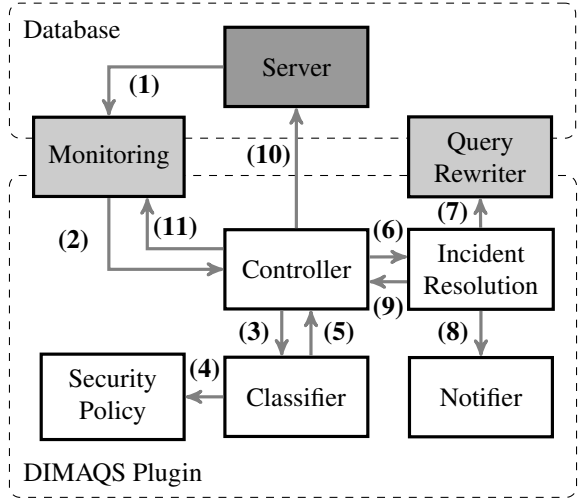


\subsection{System Architecture}\label{subsec:approach_architecture}
Figure~\ref{fig:approach_dataflow} shows the \toolAbrev system architecture. 
\toolAbrev is comprised of six components: (i)~\monitoring, (ii)~\classifier, (iii)~\policyDB, (iv)~\resolution, (v)~\notifier, (vi)~\rewriter and (vii)~\controller. The \monitoring and \rewriter components use the query parser embedded in the database server. Hence, the figure shows them as belonging to both, \toolAbrev plugin and the database server. In the following, we describe the role of every component in more detail.





\vspace{0,1cm}
\noindent\textbf{Monitoring}
\label{subsubsec:approach_monitoring}
The \monitoring component monitors all incoming queries for potentially malicious query sequences. Note that this module monitors all queries arriving through different connections, not specific to user sessions. Notifications on the occurrence of incoming queries result from the database server's audit functionality. 

\vspace{0,1cm}
\noindent\textbf{Classifier}\label{subsubsec:approach_classifier}
The \classifier component processes the incoming queries and produces a verdict whether a query is benign or malicious.
For the classification, \toolAbrev uses a CPN with our extensions.
The token colors are used to attach runtime information to the tokens, such as time-stamps, table names and modified cell values.
Since such token colors are dynamic and unbounded, conventional Petri nets would be unable to represent all the possible states. This information also provides additional information to the \toolAbrev administrator in the case of an incident\footnote{Note that \toolAbrev administrator and database administrator are different entities}.




\textit{Extensions to CPNs.} For our purposes, we extend CPNs with three new features. The first is the dynamic creation of colors for storing information inside the tokens.
The second is the ability to merge tokens that are identical except for their timestamps. This extension improves performance and does not impede classification accuracy.
The third extension allows for token expiration. Since each place in the CPN can have timeout information, this feature can be used to limit the time window of analyzed query sequences. It is highly unlikely that a malicious query sequence spawns over a long period (e.g., days), since this increases the risk of detection and complicates the attack (the database can change considerably over time). Large or absent timeouts can additionally result in a higher false positive rate since eventually all transitions might be triggered by unrelated queries. The timeout threshold is, therefore, a security parameter, which enables a trade-off between effectiveness and false alerts. In real-world attacks observed so far, attackers did not stretch malicious query sequences over long periods. Hence, even short timeouts (1-2 minutes) would work well against them. Attackers might increase the attack time window to avoid detection. However, the longer they stay connected, the higher the burden for them (since the attacks are not generally automated), and the higher the risk of being uncovered, especially given the fact that they do not know the currently used threshold parameter and, hence, have no understanding for how long they should stay connected to remain undetected.  

\vspace{0,1cm}
\noindent\textbf{Security Policy}
The \policy component holds information about patterns of malicious query sequences (or attack signatures). The CPN configuration represents it in our system -- it describes CPN's places, place actions, transitions, transition actions, transition conditions, and arcs. 

All places and transitions are named, and the arcs are each weighted with a value of 1 token.  
Each place can be assigned several place actions executed upon CPN transitions to the corresponding place. Transitions are used to check for the execution of a (next) step in a malicious query sequence. They become active when the source place contains at least one token. Each transition is assigned one transition action, representing conditions for incoming queries. For instance, they may specify the query type (e.g., query that lists tables) and the actual content of the query (such as a table name or a typical ransom message). 

A transition may also have an arbitrary number of  transition conditions which are used to evaluate the token data from the source place against the query values. Our policy includes only one transition condition, 
ensuring ransom message insertion into a previously created or modified table.



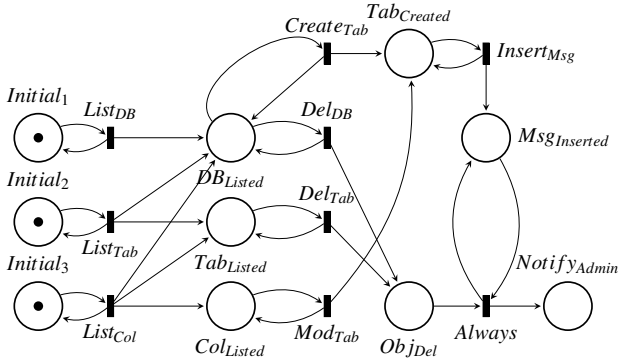
\begin{figure}
    \centering
    \scalebox{0.8}{\begin{tikzpicture}[xscale=0.8,yscale=-0.7,thin,>=stealth,
every transition/.style={fill,minimum width=1mm,minimum height=3.5mm},
every place/.style={draw,thick,minimum size=8mm}]
    \node[place,label=above:$Initial_1$,tokens=1] (i1) at (-1,2) {};
    \node[place,label=above:$Initial_2$,tokens=1] (i2) at (-1,4) {};
    \node[place,label=above:$Initial_3$,tokens=1] (i3) at (-1,6) {};
    \node[place,label=below:$DB_{Listed}$] (ol1) at (3,2) {};
    \node[place,label=below:$Tab_{Listed}$] (ol2) at (3,4) {};
    \node[place,label=below:$Col_{Listed}$] (ol3) at (3,6) {};
    \node[place,label=above:$Tab_{Created}$] (tc) at (6.7,0) {};
    \node[place,label=below:$Obj_{Del}$] (od) at (6.7,6) {};
    \node[place,label=right:$Msg_{Inserted}$] (s) at (8.3,2) {};
    \node[place,label=above:$Notify_{Admin}$] (n) at (10,6) {};
        \node[transition, label=above:$List_{DB}$] at (0.5,2) {}   
        edge [pre, bend left] (i1)
        edge [post, bend right] (i1)
        edge [post] (ol1);
    \node[transition, label=below:$List_{Tab}$] at (0.5,4) {}   
        edge [pre, bend left] (i2)
        edge [post, bend right] (i2)
        edge [post] (ol1)
        edge [post] (ol2);
    \node[transition, label=below:$List_{Col}$] at (0.5,6) {}   
        edge [pre, bend left] (i3)
        edge [post, bend right] (i3)
        edge [post] (ol1)
        edge [post] (ol2)
        edge [post] (ol3);
    \node[transition, label=above:$Create_{Tab}$] at (5,0) {}   
        edge [pre, bend left=90] (ol1)
        edge [post] (ol1)
        edge [post] (tc);
    \node[transition, label=above:$Del_{DB}$] at (5,2) {}
        edge [pre, bend left] (ol1)
        edge [post, bend right] (ol1)
        edge [post] (od);
    \node[transition, label=above:$Del_{Tab}$] at (5,4) {}   
        edge [pre, bend left] (ol2)
        edge [post, bend right] (ol2)
        edge [post] (od);
    \node[transition, label=below:$Mod_{Tab}$] at (5,6) {}   
        edge [pre, bend left] (ol3)
        edge [post, bend right] (ol3)
        edge [post, bend left=20] (tc);
    \node[transition, label=right:$Insert_{Msg}$] at (8.3,0) {}   
        edge [pre, bend left] (tc)
        edge [post, bend right] (tc)
        edge [post] (s);
    \node[transition, label=below:$Always$] at (8.3,6) {}   
        edge [pre] (od)
        edge [pre, bend left] (s)
        edge [post, bend right] (s)
        edge [post] (n);
\end{tikzpicture}}
    \caption{The CPN used to classify database transactions. All arcs are weighted with a value of 1 token.\\
    \emph{States:} ${Initial}_{x}$: initial states; ${List}_{x}$: objects listed, ${Tab}_{Created}$: table created; ${Obj}_{Del}$: object (database or table) deleted; ${MSG}_{Inserted}$: ransom message inserted; ${Notify_Admin}$: notification sent \\
    \emph{Transitions:} ${List}_{DB}$: list databases; ${List}_{Tab}$: list tables; ${List}_{Col}$ list columns; ${Create}_{Table}$: create table; ${Drop}_{Table}$: drop table; ${Modify}_{Table}$: modify table; ${Insert}_{Msg}$: insert ransom message}
    \label{fig:approach_classifierPN}
\end{figure}

We depict the CPN \add{that was tailored to the observed attacks} configured according to our security policy in Figure~\ref{fig:approach_classifierPN}.
Table~\ref{tab:approach_classifierPlaces} shows the place actions executed after putting a token on the place. 

Transitions fire when an action occurs that is specified as malicious by the \policy component. 
Note, that no single action alone is enough to transit the CPN to the "attack detected" state. Typically, the sequence of actions would be required, and their execution requires a specific order (defined by the CPN configuration) to reach the state that corresponds to attack detection.  

\add{The policy is easily adaptable to include new attack signatures by modifying the Petri net. While reconfiguration is a manual process, it is not cumbersome and can be accomplished in a reasonable amount of time\footnote{Our estimate is 30 min.}.}


\begin{table}[t]
    \centering
    \begin{tabular}{ l | l}
    \hline
    Place & Description \\ \hline
    ${DB}_{Listed}$ & Rewriting \\ 
    ${Tab}_{Listed}$ & Rewriting \\ 
    ${Col}_{Listed}$ & Rewriting \\ 
    ${Tab}_{Created}$ & Trigger creation \\ 
    ${Obj}_{Del}$ & Create backup \\
    $Notify_{Admin}$ & Create notification \\
    \hline
    \end{tabular}
    \caption{Configured actions for the places inside the CPN in Figure~\ref{fig:approach_classifierPN}. When a token reaches a place, the specified action can be executed.}
    \label{tab:approach_classifierPlaces}
\end{table}



\vspace{0,1cm}
\noindent\textbf{Incident Resolution}
When an event in the \classifier component issues an action, an action must be carried out by the \resolution module.
Possible actions are ``create backup,'' ``rewriting'' and ``create notification.''
\resolution performs the rewriting of malicious queries as well as creates backups. 

\textit{Create backup action}. Whenever the system detects a potential attack, the \resolution component will move the database, or the table dropped by an attacker to a safe place instead of deleting it. The backup copy is invisible to users (and, hence, from the attacker) so that an attacker cannot drop it again or even identify that such a backup exists. To hide backed up tables and databases from users, \resolution uses a "rewriting" action. While performing such a move, \resolution renames the protected tables to avoid name collisions.

\textit{Rewriting Action}. Rewriting actions rewrite queries to exclude tables and databases created by \toolAbrev. The \rewriter component performs these actions.

\textit{Notification action}. Notification actions are used by the \resolution component whenever there is a need to notify an administrator about a detected attack. The \notifier component performs this notification as described below.

\vspace{0,1cm}
\noindent\textbf{Notifier}
The \notifier component informs about security incidents by sending an email to the \toolAbrev administrator.
The gathered information relevant to the incident is attached to the notification so that the administrator can evaluate the incident and respond accordingly (e.g., restore the deleted table). 

\vspace{0,1cm}
\noindent\textbf{Query Rewriter}
The \rewriter component rewrites queries to exclude tables and databases created by \toolAbrev from query results.
For a `rewriting' action, the \rewriter receives the name of the table and, if applicable, the name of the database from the \resolution component.
If the queries are nested, the \rewriter extracts them into sub-queries, rewriting each sub-query separately. 
For instance, a query dropping a table will be rewritten to move the table to a safe storage space. This operation happens without any indication to the attacker. Additionally, some statements that list tables and databases will be rewritten to exclude the hidden information from query results.


\vspace{0,1cm}
\noindent\textbf{Controller}
The \Controller component connects all other \toolAbrev system segments. It is the central element that orchestrates the processing of incoming queries by other components, e.g., through invocation of the \classifier component to classify the query as malicious or benign, or the \resolution component to initiate incident resolution upon attack detection.


\subsection{Component Interaction}
\label{subsec:approach_beginMalicious}


Figure~\ref{fig:approach_dataflow} depicts the interaction between the components during query processing.
The database server first receives the query and then notifies \monitoring (1).
If \monitoring raises an alert for a potentially malicious query type, the \controller is notified (2).
The \controller then forwards the suspicious query to the \classifier (3) for evaluation.
The \classifier is configured using the security policy from the \policyDB (4) and returns the classification result to the \controller (5).
There are two possible outcomes: the query's classification is either benign or malicious. In a former case, the \controller terminates its actions, and the server executes the query as-is (10).
In the latter case, the query is considered malicious, and the \controller calls \resolution (6), which in turn backs up dropped tables and rewrites the malicious query using \rewriter (7). It then invokes the \notifier to inform the administrator about an incident (8). 
The \controller then receives the rewritten "disarmed" query from \resolution (9).
The database server then executes the query (10).
The \controller informs \monitoring when additional objects need to be observed (11), e.g., when a query creates new tables.

\section{Implementation}\label{sec:implementation}

\toolAbrev design is generic and can be applied to different database technologies. For the sake of illustration, we have chosen to prototype it for MySQL servers -- our implementation is realized as MySQL plugin compatible with MySQL server versions 5.7.x.
To function, \toolAbrev requires our own Petri net implementation library \libPetri
as well as the mysqlservices library provided by the MySQL server.
We chose the C++11 language for \toolAbrev since it is the default language for MySQL plugins. 
\toolAbrev consists of 4908 lines of code (LoC), while \libPetri results in 1008 LoC.

\subsection{Plugin Integration}

The plugin is loaded during MySQL server start-up and registers itself as an auditing plugin. 

The MySQL server plugin interface provides notifications~\cite{MySQLMan} for the following useful events:
\begin{itemize}[itemsep=0pt]
    \item \texttt{MYSQL\_AUDIT\_CONNECTION\_CLASS}, 
    \item \texttt{MYSQL\_AUDIT\_CONNECTION\_CONNECT}, 
    \item \texttt{MYSQL\_AUDIT\_CONNECTION\_DISCONNECT}, 
    \item \texttt{MYSQL\_AUDIT\_PARSE\_CLASS},  
    \item \texttt{MYSQL\_AUDIT\_PARSE\_POSTPARSE}.
\end{itemize}


Notifications of the \texttt{MYSQL\_AUDIT\_PARSE\_CLASS class} provide an event of a single to-be-executed query. Queries, however, could also be nested. 

\sloppy
Per default, the MySQL server does not provide any event that returns the atomic values of database elements affected by \texttt{INSERT}, \texttt{UPDATE}, and \texttt{DELETE} queries. 
These queries are typical for the use in attacks like mimicry, e.g., for the insertion of ransom messages.
To allow us to access the atomic values, we create triggers. We generate ``before INSERT/UPDATE'' triggers for every table. 
In these triggers, we execute a user-defined function.
This function forwards the values affected by the queries to the controller for evaluation.

As detailed in the MySQL trigger syntax~\cite{MySQLMan}, a trigger becomes associated with a table named \texttt{tbl\_name}.
This name must refer to a permanent table, which means that a trigger does not applay to a temporary table or a view.
This limitation does not affect our solution since it is unlikely that an attacker would attack data stored in temporary tables. 

\subsection{Component Implementation}

\sloppy
In the following, we detail the implementation of \toolAbrev modules. 

\vspace{0.1cm}
\noindent\textbf{Monitoring}
Additional triggers are required to access information that is not transparent to the \toolAbrev plugin when using MySQL's audit features.
Trigger creation occurs when loading the plugin, and existing triggers are recreated after server startup since the database structure might have changed. 
Trigger creation within so-called ``stored procedures'' or ``stored functions,'' the conventional concepts supported by the MySQL server is not possible. 
Due to this limitation, the creation must be within the plugin code. 
The function \texttt{dimaqs\_plugin\_init()} performs the creation of the additional triggers and is called directly after initialization of the server and before entering the listening state.
\texttt{dimaqs\_plugin\_init()} creates a trigger for every non-virtual database. 
Virtual databases are databases that contain read-only views rather than base tables and have no database files associated with them. Hence, protection of virtual databases is not necessary.

\sloppy
The \texttt{INSERT} and \texttt{UPDATE} triggers call \texttt{eval\_value()}. Several values are passed to that function, namely (1) schema name, (2) table name, and (3) new column values.
Using this structure, we can identify inserted/updated values.



\vspace{0.1cm}
\noindent\textbf{Classifier}
The \classifier is implemented using our library \libPetri. 
\libPetri is a C++ library implementing the functionality of colored Petri nets. 
It includes dynamic coloring, token timeout and token merging features mentioned above.
Since \libPetri has been developed explicitly for \toolAbrev, it carries no additional feature overhead. Thus, \libPetri contains all necessary functionality within around 1008 of LoC.

\libPetri keeps track of all active transitions. 
Since all our arcs in \classifier are weighted with the value one as seen in Figure~\ref{fig:approach_classifierPN}, active transitions have tokens on all input places. 
If the to-be-classified query matches the action attributed to an active transition, that transition fires.
When transferring a token to a place with an associated action, that action executes with the corresponding parameters. 
Until completion of these actions, the \classifier does not accept additional queries. 

\vspace{0.1cm}
\noindent\textbf{Security Policy}
The \policyDB is a database that contains tables holding the information about the actions that can fire transitions (e.g., the regular expression for detecting the ransom message) and the places with their associated actions.
\classifier processes this information on startup and during classification.

\vspace{0.1cm}
\noindent\textbf{Incident Resolution} The \Resolution backs up dropped databases and deleted values.
The renaming of databases is not trivial due to MySQL limitations. 
MySQL added a command to carry out a database renaming called '\texttt{RENAME DATABASE <database\_name>}.' 
However, this command was only active through a few minor releases before its discontinuation. 
The simplest way to rename a database is to move its tables to another database.
Each moved table requires recreation of the affected triggers. 
Table renaming follows the following schema ``\texttt{<storagespace>.<object prefix>\_<dbname>\_<tablename>\_<timestamp>}'' with \texttt{storagespace} being a preconfigured variable of \toolAbrev.
The function \texttt{renameTable()} performs this renaming. If a database drop occurs, \texttt{renameDatabase()} calls the \texttt{renameTable()} for every table.


For backup actions, a 'DROP DATABASE <db\_name>' does not require rewriting. 
However, before executing, \textit{renameTable} or \textit{renameDatabase} is executed to back up the database tables. 

\vspace{0.1cm}
\noindent\textbf{Notifier}
The \notifier sends an email with all transmitted information about the suspected attack to the administrator.
The administrator's address can be configured inside the database or in a configuration file.

\vspace{0.1cm}
\noindent\textbf{Query Rewriter}
The \rewriter rewrites a query by adding a WHERE/AND condition to hide sensitive information or rewrites it entirely, e.g., for backup operations. 

\vspace{0.1cm}
\noindent\textbf{Controller}
The \controller is implemented using the \emph{visitor} design pattern.
This visitor extracts the nested statements from inside to outside.
It then forwards each extracted query to \classifier.

 \vspace{0.2cm}

\section{Evaluation}\label{sec:evaluation}

In this section, we describe our test setup and evaluate our implementation with regards to effectiveness and performance. We conclude by discussing security considerations. 



\subsection{Test Setup} \label{subsec:testbed}



\textbf{Testbed} 
To execute performance and security tests, we use the following setup.
For the database server, we use an HPE ProLiant DL360 Gen9 server~\cite{hpedl360gen9}. 
The server is equipped with a single 8-core Haswell generation Xeon E5-2640 CPU with a base clock of 2.60 GHz and a turbo clock of 3,40 GHz and packaged with a total of 20 MB of cache~\cite{intele52640v3}.
Simultaneous multithreading is enabled allowing the execution of 16 threads in parallel.
The server features 32 GB of DDR4 RAM at 2133 MHz with dual channel capability.
A 500 GB 3.5-inch hard drive provides storage I/O turning at 7.200 rpm.

For the operating system, we chose Ubuntu 16.04.4 LTS running Linux kernel 4.4.0-121. 
To provide a DBMS to evaluate against we install and run MySQL server 5.7.22 on this server.

All tests are executed directly on this server. 
Thus, the network is not a limiting factor for the benchmarks. 
Due to the performance of the server, the resources consumed by the client running in parallel to the server are expected to be negligible, and their performance influence is therefore not evaluated in this work.

\vspace{0.1cm}
\noindent \textbf{Data Sets}
We employ three data sets during our evaluation. 
The first set (malicious set) includes malicious query sequences, which we generated ourselves using information about real-world attacks collected at~\cite{GuardiCore17}. 
Our resulting query set contains query sequence permutations with an expected malicious classification, as well as their possible permutations (since an attacker may execute them in an arbitrary order). 
The full test set contains 13\,485 tests.
Each test contains nine queries. 
The first five queries of each test are to set up two databases and a table at the beginning of the experiment and remove them at the end.
Relevant to the detection are four queries: (i) listing all databases, (ii) creating a table, (iii) inserting a ransom message into this table, and (iv) dropping a table or database.
Therefore, the set performs 53\,940 queries in total.

The second set (\emph{Bibspace set}) is from the publication management system \emph{Bibspace}~\cite{bibspace}, which was gathered over 40 days from 13th of April 2018 to 22nd of May 2018 and contains a total of 52\,085 queries. 
Among them, 24\,430 are \texttt{CREATE\_TABLE\_IF\_NOT\_EXISTS} queries, 8\,357 \texttt{INSERT} queries, and 38 \texttt{DROP\_TABLE\_IF\_EXISTS} queries. 

The third query set (\emph{MediaWiki set}) is from a locally run \emph{MediaWiki}~\cite{mediawiki} with the \emph{Semantic MediaWiki}~\cite{semanticWiki} plugin enabled, collected for 50 days from 3rd of April 2018 to 22nd of May 2018. Containing 2\,514\,764 queries, it includes 69\,261 \texttt{INSERT} statements, 29\,830 \texttt{CREATE\_TEMPORARY\_TABLE} statements, and 29\,797 \texttt{DROP\_TEMPORARY\_TABLE} statements. 

We will publish the data sets along with the paper, to allow third parties to reproduce our tests and to enable follow up works to compare with our results.

\subsection{Effectiveness}\label{subsec:effectiveness}

\begin{table*}
    \centering
    \small
    \setlength\tabcolsep{4.5pt}
    \begin{tabular}{ l | c | c | c | c | c | c | c | c | c}
    Query set & $Initial_1$ & $Initial_2$ & $Initial_3$ & $DB_{Listed}$ & $Tab_{Listed}$ & $Col_{Listed}$ & $Tab_{Created}$ & $Object_{Deleted}$ & $Notify_{Admin}$ \\ \hline
    Bibspace  & 1  & 1  & 1  & 2   & 2   & 0   & 24  & 0  & 0 \\
    MediaWiki & 1  & 1  & 1  & 7   & 5   & 1   & 0  & 0  & 0 \\
    \end{tabular}
    \setlength\tabcolsep{6pt}
    \caption{Petri net state after execution of query sets}
    \label{tab:evalutation_petri_state}
    \vspace{-10pt}
\end{table*}

In the following, we evaluate the precision of the classifier module. Thus, we evaluate whether a wrongful classification of benign queries as malicious (false positives) or malicious query sequences as benign (false negatives) occurs.

\vspace{0.1cm}
\noindent\textbf{Security Policy:}
The execution policy for the \classifier is as described in Section~\ref{subsubsec:approach_classifier}.
Our policy is quite generic in the sense that we do not look for specific table or database names, but instead detect the removal or renaming of any table or database. However, we are looking for a specific pattern of the ransom message. We search for the occurrence of a BTC or Bitcoin string inside the inserted message since attackers until now requested ransom in Bitcoins\footnote{Our policy can be trivially extended to detect ransom messages requesting payments in other cryptocurrencies.}.
We used the regular expression '(\textbackslash d*[.])\{0,1\}\textbackslash d+\textbackslash s*(BTC|Bitcoin)' (case insensitive). The matching expressions are, e.g., 5 BTC|Bitcoin, .5 BTC|Bitcoin, 20.1 btc|Bitcoin.


\vspace{0.1cm}
\noindent\textbf{False Negatives:}
To test for false negatives, we used the \emph{attack set} described in Section~\ref{subsec:testbed}.
After processing all the queries from the data set by our CPN, we achieved 100\% attack detection rate and received no false negative result. This result confirms that our CPN correctly models each attack from our malicious data set. 

\vspace{0.1cm}
\noindent\textbf{False Positives:}
To test for false positives, we choose to use the \emph{Bibspace set} and the \emph{MediaWiki set}. 
The sets contain a total of 2\,566\,849 benign queries.
The \classifier performs classification of every set.
Afterward, the \classifier state shows, if \toolAbrev wrongfully detected attacks and how many false detections occurred.
If tokens reach place $N$ in \classifier, their number represents raised alerts.
For this evaluation, we disable the token timeout, to increase the potential for false positives.

Table~\ref{tab:evalutation_petri_state} shows the population of the CPN after running all the queries from the \emph{Bibspace set} through \classifier.
No token has reached the state $N$, that would have triggered an alert to the administrator.
Next, the \classifier processed the queries of the \emph{MediaWiki set}. 
Table \ref{tab:evalutation_petri_state} shows the state of CPN from Figure~\ref{fig:approach_classifierPN} after classification. 
Again, no token has reached the state $N$, and no ransom attack was detected, which is a favorable result. 

\subsection{Performance Evaluation}\label{subsec:performance}

\begin{figure}[t]
    \centering
    \begin{tikzpicture}
\begin{axis}[
width=0.48\textwidth, height=6cm,
 scaled ticks=false, tick label style={/pgf/number format/fixed},
 ybar, 
 nodes near coords,
 xmin = 0.5, xmax = 3.5,
 ymin = 0, ymax = 120,
 ytick = {0, 20, 40, 60, 80, 100},
 ylabel = throughput in \%,
 xtick = {1, 2, 3},
 xticklabels = {disabled, initialized, all active},
 xlabel = {plugin state},
 bar width=16pt,
 grid = none,
 legend style={cells={anchor=right, fill}, nodes={inner sep=1, below=-1.1ex}, at={(0.95,0.15)}, anchor=east}, area legend]
  \addplot[color = negro, fill = negro, 
  error bars/.cd,
  y dir=both,
  y explicit,
  error bar style={line width=1.5pt, gris, xshift=0mm},
  error mark options = {
  rotate = 90, 
  line width=0pt, 
  mark size = 8pt, 
  gris,
  }
  ]
    coordinates{(1,100) +- (0,0.1)
                (2,95.3) +- (0,0.1)
                (3,95.4) +- (0,0.0)};
  \addplot [color = rojo, fill = rojo,
   error bars/.cd,
  y dir=both,
  y explicit,
  error bar style={line width=1.5pt, rojol, xshift=0mm},
  error mark options = {
  rotate = 90, 
  line width=0pt, 
  mark size = 8pt, 
  rojol}
  ]
    coordinates{(1,100) +- (0,0.1)
                (2,98.2) +- (0,0.1)
                (3,96.1) +- (0,0.03)};
 \addlegendentry{sysbench}
 \addlegendentry{MediaWiki}
 \end{axis}
\end{tikzpicture}
    \caption{Performance influence of \toolAbrev for sysbench and MediaWiki. Values are normalized to the respective value for the disabled plugin.}
    \label{fig:eval_performance}
\end{figure}
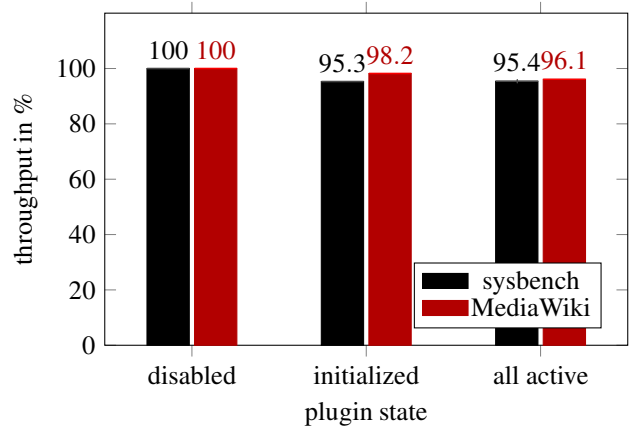

\begin{table}[t]
    \centering
    \small
    \begin{tabular}{ l | r r r | r r r|}
    Test & \multicolumn{3}{c|}{Transactions} & \multicolumn{3}{c|}{relative to} \\
    & \multicolumn{3}{c|}{per second} & \multicolumn{3}{c|}{baseline [\%]} \\
    & mean & stdev & conf& mean & stdev & conf\\
    &  &  & int & & & int \\
    \hline
    \textbf{sysbench} & & & & & & \\
    disabled & 9\,245 & 28 & $\pm 9$ & 100.0 & 0.3 & $\pm 0.1$ \\
    initialized & 8\,806 & 30 & $\pm 11$ & 95.3 & 0.3 & $\pm 0.1$ \\
    full & 8\,823 & 19 & $\pm 7$ & 95.4 & 0.2 & $\pm 0.0$ \\
    \textbf{MediaWiki} & & & & & & \\
    disabled & 2\,008 & 5 & $\pm 2$ & 100 & 0.2 & $\pm 0.1$ \\
    initialize & 1\,971 & 7 & $\pm 2$ & 98.2 & 0.3 & $\pm 0.1$ \\
    full & 1\,930 & 6 & $ \pm 13$ & 96.1 & 2.9 & $\pm 0.3$   
    \end{tabular}
    \caption{Performance without the plugin, with the plugin enabled, and with tokens in each Petri net state.}
    \label{tab:performance_transactions}
    \vspace{-12pt}
\end{table}

To evaluate the performance of the \toolAbrev plugin, we used two data sets: The \emph{MediaWiki set} described in Section~\ref{subsec:testbed} and the synthetic benchmark sysbench~\cite{sysbench}. We use sysbench 0.4.12 with 16 active threads. 
We performed three performance benchmarks:
 (1) without the plugin as a baseline measure, 
 (2) operating on a newly initialized Petri net, and
 (3) with a fully occupied Petri net with tokens in each state.
Sysbench benchmarks were run for 60 seconds per iteration, while the \emph{MediaWiki set} was classified entirely every time.
We performed every benchmark for over 50 iterations.
Table~\ref{tab:performance_transactions} shows the resulting measurements (database transactions per second). 
We report average values with standard deviation and confidence intervals (5\% quantile according to the Student's t-distribution). 
Figure~\ref{fig:eval_performance} visualizes these results.

The results show that the usage of the \toolAbrev plugin results in performance degradation of about 5~\% for sysbench.
There is no substantial difference whether the Petri net is only initialized or entirely populated (overlapping confidence intervals).
This marginal difference suggests that the overhead is not a result of querying the Petri net, but from analyzing and parsing the queries themselves.
For the \emph{MediaWiki set} performance degradation is about 2\% for the initialized Petri net and 4\% for an entirely populated net. 
This time, the influence of the set population has a more significant impact.

Our proof-of-concept prototype is not yet optimized for performance. 
Neither \toolAbrev nor \libPetri has received extensive profiling for potential bottlenecks.
Also, no compiler optimizations were enabled. Thus, performance improvements are likely possible. 


\subsection{Security Considerations}\label{subsec:securitycons}


In the following, we discuss potential attack scenarios against \toolAbrev itself and show, how our system defends itself against them. 

\vspace{0.1cm}
\noindent
\textbf{DIMAQS disabling:}
An attacker may try to disable \toolAbrev to avoid detection.   
However, such a  scenario would not be successful, since administrative privileges to the database are insufficient to perform this task. One would need to have administrative privileges to the file system to manipulate corresponding config files. As an additional burden, it is also non-trivial for an attacker to detect that the system runs under \toolAbrev observation because the \rewriter component of \toolAbrev rewrites the queries in such a way that it excludes information about \toolAbrev from the results. 

\vspace{0.1cm}
\noindent
\textbf{DIMAQS triggers removal:} A next possible attack vector is specific to MySQL implementation, which uses triggers. An attacker may attempt to delete triggers, which are used  to deliver additional information to the \toolAbrev plugin. 



To defend against this attack vector, \toolAbrev detects the removal of \toolAbrev-specific triggers. Their absence becomes obvious, whenever the plugin does not receive information about atomic values affected by the queries. Upon detection, \toolAbrev generates a notification for the \toolAbrev administrator and backups all the databases and tables affected by subsequent queries. 

\vspace{-6pt}
\section{Related Work}\label{sec:related}

In this section, we provide an overview of the related work in three domains: (i)~intrusion detection for databases, (ii)~ransomware detection, and (iii)~application of Petri Nets for intrusion detection in various application domains. 


    %
\vspace{0.1cm}
\noindent \textbf{Intrusion Detection for Databases}\label{sec:rel-databases}
There is a plethora of previous works on intrusion detection systems in databases, but none of them explicitly focused on detection of ransomware so far. 
The first line of works in this category concentrate on detection of SQL injections. Fonseca et al.~\cite{fonseca2007detecting} and Kemalis et al.~\cite{kemalis2008sql} detect anomalies in SQL commands given a training set of known valid query structures or their specifications. 
Buehrer et al.~\cite{buehrer2005using} and  Bockermann et al. \cite{bockermann2009learning} use tree structure when parsing SQL statements and then dynamically compare them with the intended queries. 
AMNESIA~\cite{halfond2005amnesia,halfond2006preventing} checks the application code for SQL queries generating automata for each query to match against dynamic requests during operation. SQLCheck~\cite{su2006essence} validates queries by adding a key at the beginning and the end of each user's input and validate syntactic correctness of  the "augmented" queries  at runtime.
In contrast to our work, all these approaches concentrate on the analysis of single queries, while we aim at the detection of malicious query sequences. 



Intrusion detection frameworks~\cite{Bertino05,chung2000demids,valeur2005learning} analyze database audit logs to detect anomalous queries by matching against role profiles.
In contrast to our work, their analysis concentrates on irregular access patterns of single SQL queries. Moreover, their analysis is bound to user profiles, while \toolAbrev performs global monitoring across user sessions. 

DAIS~\cite{liu2001dais} and the solution by Liu et al.~\cite{liu2002architectures} combine intrusion detection with the dynamic isolation of malicious and suspicious activities through rewriting of SQL statements. As a result, potentially malicious  modifications are performed on a shadowed incremental copy of the database. 
In our work, we use a similar approach to preserve copies of the values affected by potentially malicious queries.



The most similar work to ours is by Hu et al.~\cite{hu2003ident, hu2004data}, who proposed an intrusion detection system for databases using (uncolored) Petri Nets. However, Hu et al. choose to model data dependency relationships and regular data update patterns and then detect anomalies, while we model malicious query sequences and compare the sequences captured at runtime with the derived model. As such, their system requires knowledge about the legitimate state of the system, while our approach represents a signature-based misuse detection system and needs knowledge about attack patterns. As a result, our solution applies to databases of arbitrary complexity and without the need to learn about underlying data structure (which can be complex), while the solution by Hu et al. requires a training phase to gain knowledge about the database under protection. On a positive side, their approach is likely to detect previously unseen malware. The feasibility of the approach by Hu et al. however was not practically verified, since authors concentrated on theoretical aspects and did not provide any implementation and evaluation. Their concept also relies on several assumptions that simplify the model but might be too restrictive in practical scenarios. For instance, they assume low database load and that users only update the database through a limited number of fixed transactions modifying the same data items. Our solution, in contrast, operates on databases of arbitrary complexity and with good performance.

Lee et al.~\cite{lee2000intrusion} target real-time databases with regular access patterns, which occur, e.g., in data collection from sensors. They use time signatures to capture expectations about update rates and flag unexpected and possibly malicious operations. 
DIWeBa~\cite{roichman2008diweba} is an anomaly-based intrusion classifier for web databases that works at the session level by fingerprinting user sessions. 
DIDAFIT~\cite{low2002didafit} models benign query sequences and maps them to a directed graph, where graph vertices represent query signatures. Enforcement of sequence orders on the graph prevents anomalous queries.
In contrast to our work, solutions above require a training phase to learn the benign behavior of users, manual setup and knowledge of the database content, or to construct graphs of benign queries.  

Mathew et al.\cite{sunu2010data} argue that query classification based on syntax is more error-prone than observing the accessed data points since syntactically similar queries can produce significantly different results. Their system is another example of observing anomalous database access patterns, which need a training phase or some predetermined knowledge of acceptable behavior.

It is also possible to perform intrusion detection through complex event processing (CEP)~\cite{Luckham98complexevent}. Romano et al.~\cite{romano2011generic} propose a generic framework for intrusion detection through CEP, where they examine different intrusions, including policy violations, buffer overflows and SQL injections. It should be noted that CEP is not a single algorithmic concept, but rather the more general idea to infer not directly observable events from multiple, related events. In a way, our implementation with CPNs acts similarly, observing individual queries that together form a ransomware attack. On the other hand, CEP systems are mostly merely a monitoring and information processing tool, while our solution includes active components, such as the automatic table backup functionality.  

Commercial solutions, such as IBM Guardium~\cite{IBMGuardium} and IMPERVA SecureSphere~\cite{Imperva}, offer intrusion detection for databases for detection of misbehaving users. 
While detailed evaluation of these products is impossible due to their proprietary nature, we speculate that an attacker could easily evade their detection, since their analysis is bound to user sessions. 


\vspace{0.1cm}
\noindent \textbf{Ransomware Detection}
Several solutions have been proposed to detect and prevent ransomware at the file level. CryptoDrop~\cite{scaife2016cryptolock}, ShieldFS~\cite{continella2016shieldfs,continella2017shieldfs} and Redemption~\cite{kharraz2017redemption} all monitor the file system to detect intrinsic ransomware behavior, such as file type changes, file entropy, and file similarity. They differ by their choice of observed properties, and by the mechanisms provided to prevent data loss, such as providing shadowed copies of files to possibly malicious processes.
UNVEIL~\cite{kharraz2016unveil} tries to detect evasive ransomware by generating artificial user environments for dynamic analysis. 
However, their approach does not apply to server-side database ransomware.
PayBreak~\cite{kolodenker2017pay} observes the use of symmetric keys commonly used by ransomware to encrypt files and holds them in escrow. This observation enables the recovery of the decryption keys upon ransomware detection.
For the observed attacks on databases this approach hardly applicable since the files were deleted instead of encrypted.

FlashGuard~\cite{huang2017flash} and RWGuard~\cite{mehnaz2018rwguard} propose ransomware-tolerant Solid-State Drives (SSDs) which are based on the property of SSDs to perform out-of-place writes in order to mitigate long erase latency. Both operate on the firmware level and are effective in recovering encrypted files without impacting performance or lifetime.

The related work presented in this section targets client-side crypto-ransomware and is not applicable for detection of wipers at databases, as those do not use crypto primitives and do not access the file system directly. 


\vspace{0.1cm}
\noindent \textbf{Petri Nets and State Analysis}
Previous work has explored the concept of state analysis and more specifically the use of Petri nets for intrusion detection.
Kumar et al.~\cite{kumar1994pattern, kumar1999software} present a generic model and a misuse detection system for OS kernel audit logs using CPNs. This work is conceptually comparable to our work regarding the use of a Petri net to match attack patterns but focuses on intrusions in UNIX systems.  
Ilgun et al.~\cite{ilgun1995state} also focus on UNIX systems and use states and transitions to identify the necessary steps for penetrations, resulting in a flexible rule-based system to detect intrusions. Similarly, Shieh et al.~\cite{shieh1997pattern} propose a pattern-oriented model with system states and transitions to identify context-dependent patterns of intrusion. USTAT~\cite{Ilgun} is a similar state transition analysis tool for UNIX systems, which describes penetrations as sequences of state changes and uses rule-based analysis of audit trails to identify intrusions. Ho et al.~\cite{ho1998planning} describe the use of Petri nets for intrusion detection through the example of privilege escalation, again in UNIX systems. 
Helmer et al.~\cite{helmer2007software} describe a general approach using Software Fault Trees to create CPNs for intrusion detection. The work focuses on modeling of intrusions and concentrates on the detection of FTP bounce attacks.

Overall, all the works discussed above are intended for intrusion detection in other environments, mostly in UNIX systems, and are not explicitly aimed at anomaly detection in databases or for ransomware detection.

\vspace{-12pt}
\section{Conclusion and Future Work}\label{sec:conclusion}
\vspace{-6pt}

Ransomware attacks are an emerging threat, and their server-side variance that appeared recently imposes a significant threat to databases and stored data. 
In this work, we present \toolLong, the first solution against server-side ransomware.  In its heart, \toolAbrev has colored Petri nets (CPN)-based classifier, which models malicious query sequences and matches them against query sequences captured at runtime. We introduce several novel extensions for the CPN, which allow us to reduce the complexity of the system representation and achieve better performance. 

Our solution is implemented for MySQL servers and realized as a MySQL plugin, which is easily installable on existing servers. 
We evaluated our solution with regards to the precision of the attack detection as well as its performance and report no false positives, no false negatives and performance overhead under 5\% for our non-optimized implementation.




In our future work, we plan to extend \toolAbrev for detection of other attack types, since generally the framework can be used for detection of arbitrary malicious query sequences and thus not necessarily limited to ransomware detection. 
Moreover, we will investigate possibilities for automated policy generation, which is potentially achievable given more elaborate malicious data sets and by applying machine learning techniques. 
Furthermore, we plan to perform performance optimization to  decrease the imposed overhead further. 
Finally, we plan to develop new prototypes that target other database technologies\footnote{E.g., for Prolog databases the ransom message insertion and table deletion could be mapped to the \texttt{assert} and the \texttt{retractall} commands.}.

\setlength{\emergencystretch}{8em}
\bibliographystyle{plainurl}
\bibliography{bibliography}
\end{document}